\documentstyle[preprint,eqsecnum,aps]{revtex}
\begin{document}
\draft
\title{Quantum Key Distribution using Two Coherent States of Light \\
and their Superposition}
\author{Yoshihiro Nambu, Akihisa Tomita, Yoshie Chiba-Kohno, and Kazuo Nakamura}
\address{Fundamental Research Laboratories, NEC Corporation, \\
34 Miyukigaoka, Tsukuba, Ibaraki 305-8501, Japan}
\date{\today}
\maketitle

\begin{abstract}
Quantum mechanical complementarity ensures the security of the
key-distribution scheme reported by Brassard and Bennet in 1984 (BB84), but
does not prohibit use of multi-photons as a signal carrier. We describe a
novel BB84 scheme in which two nearly orthogonal coherent states carry the
key, and the superposition of these states (cat states) protects the
communication channel from eavesdropping. Information leakage to
eavesdroppers can be determined from the visibility of the interferential
fringes in the distribution of the outcome when a certain quadrature
component is measured through homodyne detection. The effect of channel loss
and detector inefficiency is discussed.
\end{abstract}


%\twocolumn
\narrowtext

\baselineskip5mm

\preprint{20000223}

\section{INTRODUCTION}

\label{SEC1}

The quantum key distribution (QKD) protocol provides a way for two remote
parties (traditionally known as Alice and Bob) to share a secure random key
by communicating over an open channel\cite{Wiesner,BB84,E92,BBM92,B92}.
Alice and Bob publicly communicate over a quantum channel and then exchange
messages over a classical channel that can be monitored but not tampered
with by an eavesdropper (Eve). Quantum mechanics ensures that any activities
of potential eavesdroppers can be detected. Even if some eavesdropping is
found, Alice and Bob can further process the obtained key (the raw key) to
extract a safe but much shorter key (the final key) by using a classical
method of error correction (a reconciling protocol) and private
amplification \cite{PR-AMP,PR-AMP2}. A secure message of equal length to the
final key can be transmitted over the classical channel by conventional
encryption methods such as the one-time pad method\cite{Vernum}. The
security of the encrypted communication depends directly on the security of
the final key.

Among the protocols proposed so far, the four-state scheme, usually referred
to as the BB84 protocol\cite{BB84}, is claimed to be provably secure under
the assumption that Alice uses a perfect single-photon source\cite{Mayers1}.
In this protocol, Alice and Bob use two conjugate bases (say, a rectilinear
basis, $+$, and a diagonal basis, $\times $) for the polarization of a
single photon. In basis $+$, they use two orthogonal states $\left|
0_{+}\right\rangle $ and $\left| 1_{+}\right\rangle $ to encode logical ``$0$%
'' and ``$1$'', respectively, and in basis $\times ,$ $\left| 0_{\times
}\right\rangle (=\left( 1/\sqrt{2}\right) \left[ \left| 0_{+}\right\rangle
+\left| 1_{+}\right\rangle \right] )$ and $\left| 1_{\times }\right\rangle
(=\left( 1/\sqrt{2}\right) \left[ \left| 0_{+}\right\rangle -\left|
1_{+}\right\rangle \right] )$. Alice transmits a random sequence of these
states through their quantum channel and Bob measures each state with a
basis randomly chosen from $\{+,\times \}$. After transmission, the basis is
revealed, which enables Bob to discard the data that Alice and Bob used a
different basis to encode and decode and that provide inconclusive results
to Bob. The remaining data, which is called the sifted key\cite{Huttner3},
should agree for Alice and Bob and yield conclusive results for Bob.

The key idea of the BB84 protocol is that simultaneous measurements of
non-commuting observables for a single quanta are forbidden by quantum
mechanical complementarity. For these non-commuting observables, the
measurement of one observable made on the eigenstate of another observable
inevitably introduces disturbance to the state because of the back reaction
of the measurement. Since Eve has no {\it a priori} information about the
randomly chosen bases of each bit in the sifted key, she is forced to guess
which observable to measure for each photon. On average, half the time Eve
will guess wrong and thus introduce a disturbance into the state. The
disturbance can be detected as a bit error by comparing parts of the sifted
key.

The theoretical QKD schemes that have been proven secure against a wide
class of attacks have involved the transmission of a single particle that is
subject to quantum mechanics. On the other hand, there has been growing
interest among researchers on quantum information processing using
multi-photon states\cite{Braunstein,Furusawa}. Several authors have extended
this idea and have recently proposed a QKD scheme that uses multi-photon
states as a quantum carrier\cite{Ralph1,Hillery,Reid}. All these authors
used squeezed states, in which the key data are encoded on continuous,
conjugate observables of the field quadrature components. Hillery further
suggested that any nonclassical field state is useful for quantum
information processing and communication \cite{Hillery}. In this paper, we
show that quantum mechanics allows use of multi-photon states as a signal
carrier in the BB84 protocol, and provide another example that supports
Hillery's suggestion by showing that a secure BB84 protocol can be
constructed by using two nearly orthogonal coherent states and the
superposition of these states (cat states).

The organization of this paper is as follows. Section \ref{SEC2} reviews the
BB84 protocol. The connection between the protocol and the information
exclusion principle proposed by Hall\cite{Hall} is discussed and a
comprehensive explanation of the principle of the BB84 protocol is given.
The importance of an exact determination of information leakage to
eavesdroppers is stressed and what is required for the BB84 protocol is
explained. Section \ref{SEC3} is devoted to the main subject of this paper.
The basic idea and the protocol of the QKD\ scheme using two coherent states
and their superposed state are presented, and the principle and security of
this scheme are discussed. Section \ref{SEC4} is mainly devoted to
discussion of the effect of channel loss and detector inefficiency for both
the present scheme and the conventional scheme. In Sec. \ref{SEC5}, we
summarize the main results of the paper.

\section{BB84 PROTOCOL}

\label{SEC2}

The BB84 protocol can most clearly be understood in terms of the information
exclusion principle\cite{Hall}. This principle provides an
information-theoretic description of quantum complementarity and imposes an
upper bound on the sum of the information gain obtained from observation of
complementary observables in a quantum ensemble. Consider two observables $A$
and $B$ of a quantum system with an $N$-dimensional Hilbert space. They are
said to be complementary if their eigenvalues are nondegenerate, and the
overlap of any two normalized eigenvectors $\left| a_{j}\right\rangle $ of $A
$ and $\left| b_{j}\right\rangle $ of $B$ satisfy $\left| \left\langle
a_{i}|b_{j}\right\rangle \right| =1/\sqrt{N}$; therefore, the eigenstates of 
$A$ are equally weighted superpositions of the eigenstates of $B$, and vice
versa. Thus, when the system is in an eigenstate of $A$, all possible
outcomes of a measurement of $B$ are equally probable; i.e., precise
knowledge of the measured value of one observable implies maximal
uncertainty of the measured value of the other. In such a case, an operator $%
B$ is the generator of shifts in the eigenvalue of any eigenstate of $A$; $%
\exp (iBl)\left| a_{j}\right\rangle =\left| a_{(j+l)%
%TCIMACRO{\func{mod}}%
%BeginExpansion
\mathop{\rm mod}%
%EndExpansion
N}\right\rangle $, and vice versa, $\exp (iAm)\left| b_{j}\right\rangle
=\left| b_{(j-m)%
%TCIMACRO{\func{mod}}%
%BeginExpansion
\mathop{\rm mod}%
%EndExpansion
N}\right\rangle $\cite{Kraus,Wootters2}.

Hall proved an inequality concerning information gain obtained by the
measurement of complementary observables $A$ and $B$ on a system in
arbitrary state $\rho $. Let $\rho $ be a state of an given ensemble which
is prepared with {\it a priori} probability $p_{i}$ in the known state $\rho
_{i}$, so $\rho =\sum_{i}p_{i}\rho _{i}$. The initial entropy of the system
is $H_{int}=H(\rho )=-\sum_{i}p_{i}\log _{2}p_{i}$ (in bits). Given the
conditional probability $P(a_{j}|\rho _{i})=tr(\rho _{i}A_{j})$ for
obtaining outcome $a_{j}$ when measuring an observable $A$ of the state
prepared in $\rho _{i}$, where $A_{j}=\left| a_{j}\right\rangle \left\langle
a_{j}\right| $, we can compute the {\it a posteriori} probability $Q(\rho
_{i}|a_{j})$ for preparation $\rho _{i}$ by Bayes's theorem as $Q(\rho
_{i}|a_{j})=P(a_{j}|\rho _{i})p_{i}/q_{j}$, where $q_{j}=\sum_{i}P(a_{j}|%
\rho _{i})p_{i}$ is the {\it a priori} probability for the occurrence of
outcome $a_{j}$. After the measurement, the average entropy (in bits)
becomes $H_{fin}=H(\rho |A)=-\sum_{j}q_{j}\sum_{i}Q(\rho _{i}|a_{j})\log
_{2}Q(\rho _{i}|a_{j})$. The average information gain (in bits) is $I(\rho
;A)\equiv H_{ini}-H_{fin}=H(\rho )-H(\rho |A)=-\sum_{i}p_{i}\log
_{2}p_{i}+\sum_{j}q_{j}\sum_{i}Q(\rho _{i}|a_{j})\log _{2}Q(\rho _{i}|a_{j})$%
;\cite{Peres1,Caves} this is also called the Shannon mutual information.
Hall proved that the inequality

\begin{equation}
I\left( \rho ;A\right) +I\left( \rho ;B\right) \leq 2\log _{2}N\xi =\log
_{2}N  \label{InfEx}
\end{equation}
holds for the measurement of complementary observables $A$ and $B$ on a
system in arbitrary state $\rho $, where $\xi =\max \left| \left\langle
a_{j}|b_{j}\right\rangle \right| =1/\sqrt{N}$\cite{Hall}. When $N=2$,
inequality (\ref{InfEx}) means that the recoverable information can never
exceed the maximal von Neumann entropy ($S_{\max }=1)$ bit of the system,
which depends only on the dimension -- the number of distinguishable pure
states -- of the Hilbert space in which the signal states lie. Inequality (%
\ref{InfEx}) states that the information gain corresponding to the
measurement of an observable can be maximized only at the expense of the
information gains corresponding to the measurement of the complementary
observable. Hall named inequality (\ref{InfEx}) the information exclusion
principle and showed that it is closely related to Heisenberg's uncertainty
principle and Bohr's complementary principle\cite{Hall}.

To see how the information exclusion principle relates to the BB84 protocol,
let us briefly review the optimal eavesdropping strategy within an
individual-attack scheme in which each signal carrier sent by Alice is
independently subject to eavesdropping. In this strategy, Eve lets a probe
of arbitrary dimensions interact with each signal carrier independently. As
a result, each of her probes is correlated to a transmitted state and its
partial information is imprinted onto the probe. She then delays her
measurement and keeps the quantum information in her probes until she learns
the bases used by Alice and Bob from their public announcement. She finally
tries to extract as much information as possible about the transmitted
states by measuring her probes. To avoid revealing herself in too
straightforward a manner by introducing different error rates in the
different bases (because the error rate should be independent of the basis
if the errors are due to a random process), Eve applies a symmetric
eavesdropping strategy that treats the two bases on an equal footing. This
strategy has been shown to require a two-qubit probe -- i.e., a quantum
system with a four-dimensional Hilbert space\ -- and to be optimal by Fuchs 
\cite{Fuchs}. He proved that the joint unitary operation $U$ acting on the
Hilbert space of the carrier and probe is a state-dependent optimal
quantum-cloning process\cite{Buzek,Bruss,Gisin1} that is given by

\begin{eqnarray}
\left| \psi \right\rangle \left| 0_{x}\right\rangle &\rightarrow &U\left|
\psi \right\rangle \left| 0_{x}\right\rangle  \nonumber \\
&=&\sqrt{F}\left| \tilde{\psi}_{00}^{x}\right\rangle \left|
0_{x}\right\rangle +\sqrt{D}\left| \tilde{\psi}_{01}^{x}\right\rangle \left|
1_{x}\right\rangle ,  \label{Sym1} \\
\left| \psi \right\rangle \left| 1_{x}\right\rangle &\rightarrow &U\left|
\psi \right\rangle \left| 1_{x}\right\rangle  \nonumber \\
&=&\sqrt{D}\left| \tilde{\psi}_{10}^{x}\right\rangle \left|
0_{x}\right\rangle +\sqrt{F}\left| \tilde{\psi}_{11}^{x}\right\rangle \left|
1_{x}\right\rangle ,  \label{Sym2} \\
\left| \tilde{\psi}_{mn}^{x}\right\rangle &\equiv &\left\langle m_{x}\right|
U\left| \psi \right\rangle \left| n_{x}\right\rangle /\left| \left\langle
m_{x}\right| U\left| \psi \right\rangle \left| n_{x}\right\rangle \right| ,
\label{Sym3}
\end{eqnarray}
for $x=+,\times $ and $m,n=0,1$, where $F+D=1$, and $\left| \psi
\right\rangle $ is the initial state of Eve's probe and $\left| \tilde{\psi}%
_{mn}^{x}\right\rangle $ is its normalized state after interaction. The four
possible states of $\left| \tilde{\psi}_{mn}^{x}\right\rangle $ are not
necessarily orthogonal to each other, but all scalar products other than $%
\left\langle \tilde{\psi}_{11}^{x}|\tilde{\psi}_{00}^{x}\right\rangle
=\left\langle \tilde{\psi}_{00}^{x}|\tilde{\psi}_{11}^{x}\right\rangle
=\left\langle \tilde{\psi}_{10}^{x}|\tilde{\psi}_{01}^{x}\right\rangle
=\left\langle \tilde{\psi}_{01}^{x}|\tilde{\psi}_{10}^{x}\right\rangle
\equiv {\cal V}$ must be zero and ${\cal V}$ must equal $F-D$ in order to
symmetrize the strategy\cite{Cirac}.

Let us calculate the probabilities that Bob and Eve will correctly infer the
state transmitted by Alice when Eve uses this eavesdropping strategy. These
probabilities are characterized by the conditional probability $P(j|i)$ of
obtaining outcome ${\it j}${\it ,} given that state $\rho _{i}$ was
transmitted by Alice. Suppose that Alice transmits either $\rho _{0x}=\left|
0_{x}\right\rangle \left\langle 0_{x}\right| $ or $\rho _{1x}=\left|
1_{x}\right\rangle \left\langle 1_{x}\right| $. Bob's marginal density
matrices $\rho _{ix}^{B}$, and Eve's, $\rho _{ix}^{E}$, after the
signal-probe interaction and without learning each other's measurement
outcomes (nonselective measurement), are easily calculated as

\begin{eqnarray}
\rho _{0x}^{B} &=&{\cal E}^{B}(\rho _{0x})=tr_{E}U\left| \psi \right\rangle
\left\langle \psi \right| \otimes \rho _{0x}U^{-1}  \nonumber \\
&=&F\rho _{0x}+D\rho _{1x},  \label{Bob1} \\
\rho _{1x}^{B} &=&{\cal E}^{B}(\rho _{1x})=tr_{E}U\left| \psi \right\rangle
\left\langle \psi \right| \otimes \rho _{1x}U^{-1}  \nonumber \\
&=&D\rho _{0x}+F\rho _{1x},  \label{Bob2} \\
\rho _{0x}^{E} &=&{\cal E}^{E}(\rho _{0x})=tr_{B}U\left| \psi \right\rangle
\left\langle \psi \right| \otimes \rho _{0x}U^{-1}  \nonumber \\
&=&F\sigma _{00}^{x}+D\sigma _{01}^{x},  \label{Eve1} \\
\rho _{1x}^{E} &=&{\cal E}^{E}(\rho _{1x})=tr_{B}U\left| \psi \right\rangle
\left\langle \psi \right| \otimes \rho _{1x}U^{-1}  \nonumber \\
&=&D\sigma _{10}^{x}+F\sigma _{11}^{x},  \label{Eve2}
\end{eqnarray}
where $\sigma _{mn}^{x}=\left| \tilde{\psi}_{mn}^{x}\right\rangle
\left\langle \tilde{\psi}_{mn}^{x}\right| $ and ${\cal E}(\rho )$ is a
trace-preserving, completely positive, linear map of the density operators
of Alice, and Eqs. (\ref{Bob1})-(\ref{Eve2}) define the unitary
representation\cite{Schumacher1,Schumacher2,Preskill} of this map. When Bob
performs a standard measurement on the sifted key, the conditional
probabilities of Bob's inference of his signal {\it j} when Alice sends
signal {\it i }are, for $x=+,\times $ and $i,j=0,1$,

\begin{eqnarray}
P_{x}^{AB}(j|i) &=&tr\left( \rho _{ix}^{B}\left| j_{x}\right\rangle
\left\langle j_{x}\right| \right)  \nonumber \\
&=&\left\{ 
\begin{array}{ll}
F=\frac{1+{\cal V}}{2} & \mbox{if}\ i=j \\ 
D=\frac{1-{\cal V}}{2} & \mbox{if}\ i\neq j
\end{array}
\right. .  \label{Bob3}
\end{eqnarray}
On the other hand, Eve's strategy is first to distinguish between two
mutually orthogonal sets $S_{i}=\{\sigma _{i0}^{x},\sigma _{i1}^{x}\}$ $%
(i=0,1)$ that can be perfectly separated with a standard measurement. She
next performs a measurement that distinguishes between $\sigma _{00}^{x}$
and $\sigma _{11}^{x}$ or between $\sigma _{01}^{x}$ and $\sigma _{10}^{x}$,
which are not necessary mutually orthogonal ($tr\sigma _{00}^{x}\sigma
_{11}^{x}=tr\sigma _{01}^{x}\sigma _{10}^{x}\neq 0$), that gives the
smallest possible error probability. This is the best she can do in terms of
the information gained from the sifted key\cite{Cirac}. It is well known
that such a measurement is realized by standard measurement in the basis in
the Hilbert space spanned by $\left| \tilde{\psi}_{00}^{x}\right\rangle $
and $\left| \tilde{\psi}_{11}^{x}\right\rangle $ or by $\left| \tilde{\psi}%
_{01}^{x}\right\rangle $ and $\left| \tilde{\psi}_{10}^{x}\right\rangle $
that straddles these vectors\cite{Kennedy2,Davis,Huttner1,Levitin2,Helstrom}%
. This measurement gives the conditional probabilities of Eve's inference of
her signal {\it j} when Alice sends signal {\it i} as

\begin{eqnarray}
P_{x}^{AE}(j|i) &=&tr\left( \rho _{ix}^{E}\widehat{\Pi }_{j}^{x}\right) 
\nonumber \\
&=&\left\{ 
\begin{array}{ll}
\frac{1}{2}\left( 1+{\cal D}_{opt}\right) =\frac{1+\sqrt{1-{\cal V}^{2}}}{2}
& \mbox{if}\ i=j \\ 
\frac{1}{2}\left( 1-{\cal D}_{opt}\right) =\frac{1-\sqrt{1-{\cal V}^{2}}}{2}
& \mbox{if}\ i\neq j
\end{array}
\right. ,  \label{Eve3}
\end{eqnarray}
where ${\cal D}_{opt}=tr\left| \sigma _{00}^{x}-\sigma _{11}^{x}\right|
=tr\left| \sigma _{01}^{x}-\sigma _{10}^{x}\right| =\sqrt{1-{\cal V}^{2}}$
is the distance between $\sigma _{00}^{x}$ and $\sigma _{11}^{x}$ and
between $\sigma _{01}^{x}$ and $\sigma _{10}^{x}$ in the trace-class norm,
and $\widehat{\Pi }_{0}^{x}$ and $\widehat{\Pi }_{1}^{x}$ are the
projection-valued measures (PVMs) corresponding to the above detection
strategy to distinguish between $\rho _{0x}^{E}$ and $\rho _{1x}^{E}$. (Eve
also knows when Bob has received an error)\cite
{Fuchs,Helstrom,Levitin,Busch,Jaeger}. Finally, upon assuming equal {\it a
priori} probabilities $p_{0+}=p_{1+}=p_{0\times }=p_{1\times }$, Bob's
average probability ({\it a posteriori} probability) of correct (incorrect)
inference of the state transmitted by Alice, $Q_{c}^{B}$ ($Q_{e}^{B}$), is
given by $\frac{1}{2}\left( P_{+}^{AB}(j|i)+P_{\times }^{AB}(j|i)\right) $
with $i=j$ ($i\neq j$) and Eve's average probability, $Q_{c}^{E}$ ($%
Q_{e}^{E} $), is given by $\frac{1}{2}\left( P_{+}^{AE}(j|i)+P_{\times
}^{AE}(j|i)\right) $ with $i=j$ ($i\neq j$). Thus, $Q_{c}^{B}=\frac{1+{\cal V%
}}{2}$ and $Q_{c}^{E}=\frac{1+\sqrt{1-{\cal V}^{2}}}{2}$ gives Bob's and
Eve's fidelity, respectively, and $Q_{e}^{B}=\frac{1-{\cal V}}{2}$ and $%
Q_{e}^{E}=\frac{1-\sqrt{1-{\cal V}^{2}}}{2}$ gives Bob's and Eve's error
probability, respectively. $G^{B}=Q_{c}^{B}-Q_{e}^{B}={\cal V}$ and $%
G^{E}=Q_{c}^{E}-Q_{e}^{E}={\cal D}_{opt}$ are convenient measures of Bob's
and Eve's information gain\cite{Fuchs}. Since these measures satisfy $\left(
G^{B}\right) ^{2}+\left( G^{E}\right) ^{2}={\cal D}_{opt}^{2}+{\cal V}^{2}=1$%
, there is a trade-off relation between Bob's and Eve's information gain.

From an information-theoretic point of view, the mutual information $I_{AB}$
between Alice and Bob and $I_{AE}$ between Alice and Eve concerning Alice's
message is more appropriate for evaluating Bob's and Eve's knowledge about
the sifted key. Mutual information is the measure of information
successfully transmitted from input to output. Since Alice and Bob, in
general, cannot distinguish between errors caused by eavesdroppers and
errors caused by the environment, they have to assume that all errors are
due to potential eavesdroppers. As long as Bob's error rate, $Q_{e}^{B}$, is
small, the errors can be accepted and corrected by legitimate users. As a
result, Eve can obtain some information about the transmitted data. This
information leakage to Eve can be eliminated by privacy amplification\cite
{PR-AMP,PR-AMP2} at the cost of reducing the length of the final key.
Privacy amplification requires an exact determination of the upper bound of
the information leakage to Eve. Thus, the security and robustness of the
final key totally depends on this determination. The simple criterion for
obtaining a finite length for the secure final key is still an open
question, but the inequality $I_{AB}>I_{AE}$ is believed to provide a fairly
good criterion; i.e., if the channel noise is such that $I_{AB}<I_{AE}$ for
any potential eavesdropper, then Alice and Bob should consider the
transmission channel to be unsafe. On the contrary, if $I_{AB}>I_{AE}$, they
may still be able to extract a safe but much shorter cryptographic key.
Moreover, in a classical context there is, at least in principle, a way for
Alice and Bob to exploit any positive difference, $I_{AB}-I_{AE}$, to create
a reliably secret string of key bits that has a length of about $%
I_{AB}-I_{AE}$\cite{Wyner,Csiszar,Maurer}. It is therefore important to
exactly determine the upper bound of the information leakage to Eve from a
quantity that Alice and Bob can evaluate.

In the BB84 protocol, $I_{AB}$ can be evaluated directly and $I_{AE}$ can be
determined from the error rate $Q_{e}^{B}$ that Alice and Bob can evaluate.
With equiprobable signals, they are given by $I_{AB}=1-H(Q_{e}^{B})$ and $%
I_{AE}=1-H(Q_{e}^{E})$, where $H(q)=-q\log _{2}q-(1-q)\log _{2}(1-q)$ is the
entropy function (in bits) and is a nonlinear function of $q$. Since $\left(
Q_{e}^{B}-1/2\right) ^{2}+\left( Q_{e}^{E}-1/2\right) ^{2}=\{\left(
G^{B}\right) ^{2}+\left( G^{E}\right) ^{2}\}/4=1/4$, $Q_{e}^{B}$ and $%
Q_{e}^{E}$ are mutually related. The upper plot in Fig. 1 shows $I_{AB}$, $%
I_{AE}$ and $I_{AB}+I_{AE}$ plotted against $Q_{e}^{B}$, and the lower plot
shows $G^{B}$ and $G^{E}$. From this figure, it is clear that there is a
trade-off relation between $I_{AB}$ and $I_{AE}$ as well as a trade-off
relation between $G^{B}$ and $G^{E}$. The sum $I_{AB}+I_{AE}\ $never exceeds
unity ($I_{AB}+I_{AE}\leq 1$).

The last inequality, $I_{AB}+I_{AE}\leq 1$, is closely related to the
information exclusion principle. This is because the above eavesdropping
strategy can be alternatively viewed as a method for simultaneously
measuring non-commuting observables. To see this, consider the unitary
operation in Eqs. (\ref{Sym1}) and (\ref{Sym2}) with $x=+$, $F=1$ ($D=0$).
This operation is called measurement of intensity $\gamma $, where $%
\left\langle \tilde{\psi}_{11}^{+}|\tilde{\psi}_{00}^{+}\right\rangle
=\left\langle \tilde{\psi}_{00}^{+}|\tilde{\psi}_{11}^{+}\right\rangle =\cos
\gamma ={\cal V}$.\cite{Gisin1} When Alice and Bob have chosen the basis $+$%
, Eve causes no disturbance and obtains information about the bit to the
extent that she can distinguish the two vectors $\left| \tilde{\psi}%
_{00}^{+}\right\rangle $ and $\left| \tilde{\psi}_{11}^{+}\right\rangle $,
whose error probability is $\frac{1-\sqrt{1-{\cal V}^{2}}}{2}$. Conversely,
if Alice and Bob have chosen the basis $\times $, Eve learns nothing and
introduces an error with probability $\frac{1-{\cal V}}{2}$. Bob's and Eve's
information gains when Alice transmits bits with the $+$ basis are therefore 
$I_{AB}^{+}=1$ and $I_{AE}^{+}=1-H(\frac{1-\sqrt{1-{\cal V}^{2}}}{2})$, and
their information gains when Alice transmits the bits with the $\times $
basis are $I_{AB}^{\times }=1-H(\frac{1-{\cal V}}{2})$ and $I_{AE}^{\times
}=0$. Thus, this operation is asymmetric with respect to the basis used in
which Eve obtains information on the bits sent with one basis at the cost of
a disturbance in the bits sent with the other basis. In this operation, Eve
obtains information only about the observable $P_{+}(=\left|
i_{+}\right\rangle \left\langle i_{+}\right| )$ of the $+$ basis\ on the
system, while Bob obtains information about both $P_{+}$\ and $P_{\times
}(=\left| i_{\times }\right\rangle \left\langle i_{\times }\right| )$ of the 
$\times $ basis. Thus, when Bob observes $P_{\times }$ and Eve observes $%
P_{+}$, the above operation provides a method for simultaneously measuring
complementary observables, $P_{+}$ and $P_{\times }$ in which the outcome
for Eve gives the information $I_{AE}^{+}=I(\rho ;P_{+})$ and that for Bob
gives $I_{AB}^{\times }=I(\rho ;P_{\times })$ satisfying $%
I_{AE}^{+}+I_{AB}^{\times }\leq 1$.

When we extend this argument to the symmetric operation associated with an
optimal eavesdropping strategy, we find $I_{AB}^{+}=I_{AB}^{\times }\equiv
1-H(\frac{1-{\cal V}}{2})$ and $I_{AE}^{+}=I_{AE}^{\times }\equiv 1-H(\frac{%
1-\sqrt{1-{\cal V}^{2}}}{2})$ because Bob's and Eve's information gains are
independent of the basis Alice chose. We thus find that the symmetric
operation provides a method for simultaneously measuring two complementary
observables, $P_{+}$ and $P_{\times }$, even when Bob observes $P_{+}$. In
this case, the outcome for Eve gives the information $I_{AE}^{\times
}=I(\rho ;P_{\times })$ and that for Bob gives $I_{AB}^{+}=I(\rho ;P_{+})$.
When we also take into account the fact that the sifted key involves only
the data for which Alice's and Bob's bases agree, the above arguments imply
that Bob's average information gain on the sifted key is given by 
\begin{equation}
I_{AB}=\frac{1}{2}\{I\left( \rho _{i\times };P_{\times }\right) +I\left(
\rho _{i+};P_{+}\right) \},  \label{InfBob}
\end{equation}
whereas Eve's information gain is given by 
\begin{equation}
I_{AE}=\frac{1}{2}\{I\left( \rho _{i\times };P_{+}\right) +I\left( \rho
_{i+};P_{\times }\right) \}  \label{InfAlice}
\end{equation}
for the symmetric operation.

We can now see that the information exclusion principle leads to the
inequality $I_{AB}+I_{AE}\leq 1$. Since the bases Alice and Bob used in the
BB84 protocol are conjugate, $\left| \left\langle 0_{\times
}|0_{+}\right\rangle \right| =\left| \left\langle 0_{\times
}|1_{+}\right\rangle \right| =\left| \left\langle 1_{\times
}|0_{+}\right\rangle \right| =\left| \left\langle 1_{\times
}|1_{+}\right\rangle \right| =1/\sqrt{2}$ holds, and it follows from the
information exclusion principle that the inequalities

\begin{eqnarray}
I\left( \rho _{i+};P_{+}\right) +I\left( \rho _{i+};P_{\times }\right) &\leq
&1  \label{InfEx1} \\
I\left( \rho _{i\times };P_{+}\right) +I\left( \rho _{i\times };P_{\times
}\right) &\leq &1  \label{InfEx2}
\end{eqnarray}
should hold. Equations (\ref{InfBob}) and (\ref{InfAlice}) and inequalities (%
\ref{InfEx1}) and (\ref{InfEx2}) imply that $I_{AB}+I_{AE}\leq 1$. We
therefore conclude that the bound on the sum of Bob's and Eve's information $%
I_{AB}+I_{AE}\leq 1$ is a direct consequence of the information exclusion
principle; that is, the sum can never exceed the maximal amount of
information that can be encoded in a two-state system. This condition must
be met for the BB84 protocol to be secure. It is therefore essential in the
BB84 protocol to limit the size of the signal space $N$, as is easily found
from Eq. (\ref{InfEx}). Thus, the conventional BB84 protocol requires use of
a single-photon carrier with a limited degree for freedom of information
encoding such as polarization encoding. Meeting this condition ensures that
no eavesdropping strategy can break this bound. Then, we can safely say that
the information leakage to Eve $I_{AE}$ is bounded by $1-I_{AB}$ which Alice
and Bob can also evaluate from the bit error rate in Bob's data. Only in
such a case, can we establish a provably secure final key by the subsequent
privacy amplification.

It is helpful for later discussion to point out that the information
exclusion principle is directly related to the fundamental relation between
fringe visibility ${\cal V}$ and which-way information (path
distinguishability) ${\cal D}_{opt}$ in one-particle interferometry\cite
{Englert,Mandel,Durr,Jaeger2,Jaeger3,Wootters}. To demonstrate this point,
we note that the identities $\left| 0_{\times }\right\rangle \left\langle
0_{\times }\right| +\left| 1_{\times }\right\rangle \left\langle 1_{\times
}\right| =\left| 0_{+}\right\rangle \left\langle 0_{+}\right| +\left|
1_{+}\right\rangle \left\langle 1_{+}\right| \equiv I$, $\left| 0_{\times
}\right\rangle \left\langle 1_{\times }\right| +\left| 1_{\times
}\right\rangle \left\langle 0_{\times }\right| \equiv \left|
0_{+}\right\rangle \left\langle 0_{+}\right| -\left| 1_{+}\right\rangle
\left\langle 1_{+}\right| $, and $\left| 0_{+}\right\rangle \left\langle
1_{+}\right| +\left| 1_{+}\right\rangle \left\langle 0_{+}\right| \equiv
\left| 0_{\times }\right\rangle \left\langle 0_{\times }\right| -\left|
1_{\times }\right\rangle \left\langle 1_{\times }\right| $ hold for a
two-state system. We then find that Bob's marginal density matrices $\rho
_{0x}^{B}$ or $\rho _{1x}^{B}$ can be rewritten in terms of the
complementary basis as \widetext

\begin{eqnarray}
\rho _{0\times }^{B} &=&\frac{1}{2}\left\{ \left| 0_{+}\right\rangle
\left\langle 0_{+}\right| +\left| 1_{+}\right\rangle \left\langle
1_{+}\right| \right. \left. +{\cal V}\left( \left| 0_{+}\right\rangle
\left\langle 1_{+}\right| +\left| 1_{+}\right\rangle \left\langle
0_{+}\right| \right) \right\} ,  \label{Vis1} \\
\rho _{1\times }^{B} &=&\frac{1}{2}\left\{ \left| 0_{+}\right\rangle
\left\langle 0_{+}\right| +\left| 1_{+}\right\rangle \left\langle
1_{+}\right| \right. \left. -{\cal V}\left( \left| 0_{+}\right\rangle
\left\langle 1_{+}\right| +\left| 1_{+}\right\rangle \left\langle
0_{+}\right| \right) \right\} ,  \label{Vis2} \\
\rho _{0+}^{B} &=&\frac{1}{2}\left\{ \left| 0_{\times }\right\rangle
\left\langle 0_{\times }\right| +\left| 1_{\times }\right\rangle
\left\langle 1_{\times }\right| \right. \left. +{\cal V}\left( \left|
0_{\times }\right\rangle \left\langle 1_{\times }\right| +\left| 1_{\times
}\right\rangle \left\langle 0_{\times }\right| \right) \right\} ,
\label{Vis3} \\
\rho _{1+}^{B} &=&\frac{1}{2}\left\{ \left| 0_{\times }\right\rangle
\left\langle 0_{\times }\right| +\left| 1_{\times }\right\rangle
\left\langle 1_{\times }\right| \right. \left. -{\cal V}\left( \left|
0_{\times }\right\rangle \left\langle 1_{\times }\right| +\left| 1_{\times
}\right\rangle \left\langle 0_{\times }\right| \right) \right\} .
\label{Vis4}
\end{eqnarray}
\narrowtext
\noindent These equations are isomorphic to the equations describing
one-particle interferometry where ${\cal V}$ gives the fringe visibility and 
${\cal D}_{opt}=\sqrt{1-{\cal V}^{2}}$ gives the maximal which-way
information (path distinguishability), satisfying ${\cal D}^{2}+{\cal V}%
^{2}\leq {\cal D}_{opt}^{2}+{\cal V}^{2}=$ 1\cite{Englert,Mandel}. Note that
the initial states that Alice transmitted are given by setting ${\cal V}=1$
in these equations. This implies that the noise introduced by eavesdropping
reduces the coherence (the off-diagonal terms) of the initial states, and
that Bob's bit error probability $Q_{e}^{B}=\frac{1-{\cal V}}{2}$ due to
eavesdropping can also be detected by observing the fringe visibility ${\cal %
V}$ in some kinds of interferometry.

To summarize this section, the security of the BB84 protocol totally relies
on the quantum mechanical complementarity. This complementarity gives a firm
basis for exact determination of the upper bound of the information leakage
to eavesdroppers. This enables the security of the final key for the BB84
protocol to be mathematically provable.

\section{BB84 PROTOCOL USING TWO COHERENT STATES AND THEIR SUPERPOSITION}

\label{SEC3}

The quantum mechanical complementarity and use of a single-photon carrier
ensures that there is an upper bound on the information leakage to
eavesdroppers and enables us to determine this bound from the information
gain of a legitimate user. The requirement for complementarity to be valid
states that the conjugate bases must belong to the same signal space. In
other words, if {\em the mutually complementary observables and bases are
chosen within the same signal-state space}, the BB84 protocol can be
constructed. This requirement can of course be satisfied when the
polarization space of a single photon is used to encode information. For
this purpose, we require a single-photon source, which has not yet been
realized. To overcome this difficulty, a self-checking source, the validity
of which can be self-checked, has been devised by Mayers et al.\cite
{Mayers2,Mayers3}

Alternatively, many experimental implementations of BB84 have used weak
coherent pulses (WCP), rather than single photons; in these implementations,
four equiprobable states given by 
\begin{equation}
\begin{array}{l}
\left| 0_{0}^{wcp}\right\rangle =\left| \alpha \right\rangle _{1}\left|
\alpha \right\rangle _{2},\text{ }\left| 1_{0}^{wcp}\right\rangle =\left|
\alpha \right\rangle _{1}\left| -\alpha \right\rangle _{2} \\ 
\left| 0_{\pi /2}^{wcp}\right\rangle =\left| \alpha \right\rangle _{1}\left|
i\alpha \right\rangle _{2},\text{ }\left| 1_{\pi /2}^{wcp}\right\rangle
=\left| \alpha \right\rangle _{1}\left| -i\alpha \right\rangle _{2}
\end{array}
\label{Coh4}
\end{equation}
were used.\cite{Yuen,Huttner2,Mu,Townsend1,Townsend2,Townsend3} Note that $%
\left| \pm i\alpha \right\rangle =\left( e^{\mp \frac{\pi }{4}}/\sqrt{2}%
\right) \left[ \left| \alpha \right\rangle \mp i\left| -\alpha \right\rangle 
\right] +O(\alpha ^{2})$. Therefore, if we consider only the first order in $%
\alpha $ (i.e., consider only a single-photon component), the four states
would behave much like the ideal BB84 states. However, if we consider higher
orders in $\alpha $, the two states in one basis $\left|
i_{0}^{wcp}\right\rangle $ are no longer linear combinations of the two
states in the other basis $\left| i_{\pi /2}^{wcp}\right\rangle $, and thus
do not satisfy the above requirement\cite{Brassard}. As a result, this
implementation is vulnerable to eavesdropping. When $\alpha $ is large,
these states are four non-orthogonal states lying in a four-dimensional
signal state space instead of two sets of two orthogonal states lying in the
two-dimensional signal state space used in the original single-photon
implementation. There are eavesdropping strategies that make use of the
linear independence of the four states. Figure 2 illustrates the relevant
subspace of the four states in the entire Hilbert space (the Fock space).
Because of the linear independence of the states, there are non-overlapping
subspaces in the four states. The states lying in this subspace can be
perfectly distinguished from each other, and a skillful eavesdropper can
make use of this flaw to obtain information about the key {\em without
detection}\cite{Ralph1,Hillery,Reid,Brassard}.

For example, Reid has described the conclusive-measurement attack, in which
Eve can sometimes get full information by using an appropriate ``positive
operator-valued measure'' (POVM)\cite
{Peres1,Preskill,Helstrom,Davies2,Kraus2} that conclusively distinguishes
such linearly independent states\cite{Reid}. Such measurement yields no
information about the state most of the time, but it sometimes identifies
the state unambiguously. Another strategy, called the generalized
beamsplitter attack, has also been reported on by several authors \cite
{Ralph1,Yuen,Huttner2,Lo}. Since the polarization and photon number are
independent observables, there is no problem in principle in selecting a few
pulses with two or more photons and separating them into two one-photon
pulses without changing the polarization, for example, by means of quantum
nondemolition measurement\cite{Yamamoto}. Both these attacks are fatal, in
particular, if the channel loss between Alice and Bob is large enough. This
is because Eve can recreate the state near Bob and send it to him whenever
she is able to measure the signal state unambiguously and can suppress the
signal without causing errors and reducing the bit rate by substituting a
less lossy channel. As a result, Eve can obtain information about the key
seemingly without introducing errors in the transmission. In addition, most
importantly, there is no security proof for the BB84 with WCP
implementations as well as the scheme using the two nonorthogonal states\cite
{B92}. This is because the quantum mechanical complementary can not work
effectively in the WCP implementation, and there is no principle for
reliably estimating the upper bound of the information leakage to
eavesdroppers. This weakness of the WCP scheme arises because the states $%
\left| i_{0}^{wcp}\right\rangle $ and$\left| i_{\pi /2}^{wcp}\right\rangle $
are linearly independent if we consider the multi-photon components of the
signal states\cite{Brassard}. Thus, the use of the four coherent states in
Eq. (\ref{Coh4}) with a large $\alpha $ is inappropriate for the BB84
protocol, and Alice and Bob must use dim coherent pulses each of which, on
average, typically contain 0.1 photons for the WCP scheme to approximate the
single-photon scheme.

Nevertheless, quantum mechanical complementarity does not forbid us to use
the multi-photon state to implement the BB84 protocol. To see this, we
consider the scheme depicted in Fig. \ref{F3}. Two nearly orthogonal
coherent states $\left| \alpha \right\rangle $ and $\left| -\alpha
\right\rangle $ are used to carry the key and the superposition of these
states $(\left| \alpha \right\rangle \pm \left| -\alpha \right\rangle )/%
\sqrt{2(1\pm \kappa )}$ is used to prevent eavesdropping. Here, $\kappa $ is
the overlap of the two coherent states $\left| \alpha \right\rangle $ and $%
\left| -\alpha \right\rangle $; i.e., $\kappa =\left| \left\langle \alpha
\right| -\alpha \rangle \right| =e^{-2\left| \alpha \right| ^{2}}$. These
states are the ``Schr\"{o}dinger's cat states'' and are parity eigenstates
that lie within the relevant two-dimensional signal subspace spanned by $%
\left\{ \left| \alpha \right\rangle ,\left| -\alpha \right\rangle \right\} $
in the Fock space\cite{Yurke,Cochrane,Buzek2}. These four states satisfy the
requirement for quantum mechanical complementarity to be valid, and would
therefore behave much like ideal BB84 states.

In the following, we describe the protocol and explain how eavesdropping is
detected. Consider the following protocol using only three, instead of four,
states. This protocol is not the original BB84 protocol and is less
efficient, but it is enough to explain the basic idea of the present scheme.

\begin{enumerate}
\item  Alice first chooses a subset of random positions within a sequence of
data being transmitted.

\item  She then transmits random bits encoded with a set of nearly
orthogonal states $\left| 0_{+}\right\rangle =\left| \alpha \right\rangle $
and $\left| 1_{+}\right\rangle =\left| -\alpha \right\rangle $ for the
chosen subset (the first subset) which provides a raw key.

\item  She also transmits either $\left| 0_{\times }\right\rangle =(\left|
\alpha \right\rangle -\left| -\alpha \right\rangle )/\sqrt{2(1-\kappa )}$ or 
$\left| 1_{\times }\right\rangle =(\left| \alpha \right\rangle +\left|
-\alpha \right\rangle )/\sqrt{2(1+\kappa )}$ for the remaining subset (the
second subset) which will be used only to detect eavesdropping.

\item  Alice also transmits a strong local oscillator beam (LO) with its
polarization rotated so as to be orthogonal to the signal beam on the same
channel by mixing the beams on a polarizing beamsplitter. The mixed beams
are then transmitted to Bob.

\item  Bob uses a polarizing beamsplitter to separate the LO from the
channel. The polarization of the LO is rotated by $\pi /2$ using a $\lambda
/2$-plate so as to match that of the signal. With this strong LO, Bob
performs balanced homodyne detection\cite{Yuen2} to measure the single
field-quadrature $\hat{X}(\theta )=\hat{x}_{a}\cos \theta +\hat{p}_{a}\sin
\theta =(1/\sqrt{2})[e^{-i\theta }\hat{a}+e^{i\theta }\hat{a}^{\dagger }]$
of the signal when he receives it, where $\hat{x}_{a}=(1/\sqrt{2})[\hat{a}+%
\hat{a}^{\dagger }]$, $\hat{p}_{a}=(1/\sqrt{2}i)[\hat{a}-\hat{a}^{\dagger }]$%
. If we assume that $\alpha $ is real for simplicity, then $\theta $ is the
advance of the signal phase relative to the LO phase (which is Bob's
controllable parameter). He randomly varies $\theta $ between $0$ and $\pi
/2 $ by changing the LO phase with phase shifter A. (It is possible for
Alice and Bob to calibrate the phase $\theta $ without introducing any
vulnerability.)

\item  After transmission, Alice publicly announces the positions of the
first and second data subsets. Alice and Bob then discard the part of the
first subset of data for which Bob measured $\hat{p}_{a}$ ($\theta =\pi /2$)
and the part of the second subset of data for which he measured $\hat{x}_{a}$
($\theta =0$). Bob can obtain the sifted key from the first subset of the
remaining data.
\end{enumerate}

For the moment, let us consider a perfect detector with unit efficiency and
negligible channel loss. The effects of the detection efficiency and channel
loss are considered in the next section. In terms of the sifted key, the
conditional probability distributions $p_{i+}(x_{a})$ of Bob's output {\it x}
when Alice transmits signal {\it i }obey the Gaussian distributions:

\begin{eqnarray}
p_{0+}(x_{a}) &=&Tr\left| 0_{+}\right\rangle \left\langle 0_{+}\right|
x_{a}\rangle \left\langle x_{a}\right|  \nonumber \\
&=&\frac{1}{\pi ^{1/2}}\exp \left[ -\left( x_{a}-\left\langle \alpha
\right\rangle \right) ^{2}\right] , \\
p_{1+}(x_{a}) &=&Tr\left| 1_{+}\right\rangle \left\langle 1_{+}\right|
x_{a}\rangle \left\langle x_{a}\right|  \nonumber \\
&=&\frac{1}{\pi ^{1/2}}\exp \left[ -\left( x_{a}+\left\langle \alpha
\right\rangle \right) ^{2}\right] ,
\end{eqnarray}
where $\left\langle \alpha \right\rangle =\sqrt{2}\left| \alpha \right| $.
The standard strategy for Bob to correctly infer the state transmitted by
Alice is to set the decision threshold at $x_{a}=0$; i.e., he sets the bit
value to 0 when he obtains $x_{a}\geq 0$ and to 1 when he obtains $x_{a}<0$.
Then, his average error probability has finite value $Q_{e}^{B}(\alpha )=%
\frac{1}{2}%
%TCIMACRO{\func{Erfc}}%
%BeginExpansion
\mathop{\rm Erfc}%
%EndExpansion
\left[ \left\langle \alpha \right\rangle \right] $, where $%
%TCIMACRO{\func{Erfc}}%
%BeginExpansion
\mathop{\rm Erfc}%
%EndExpansion
[x]$ is the complementary error function defined by $%
%TCIMACRO{\func{Erfc}}%
%BeginExpansion
\mathop{\rm Erfc}%
%EndExpansion
[x]\equiv \left( 1/\sqrt{2\pi }\right) \int_{x}^{\infty }\exp \left[ -\tau
^{2}\right] d\tau $\cite{Osaki}. This is because the two coherent states $%
\left| \alpha \right\rangle $ and $\left| -\alpha \right\rangle $ are not
orthogonal. Bob also checks the second subset of remaining data to detect
possible eavesdropping. Provided that Alice transmits the $\left| 1_{\times
}\right\rangle $ state for the second subset, the associated conditional
probability distribution $p_{1\times }(p_{a})$ is \widetext

\begin{equation}
p_{1\times }(p_{a})=Tr\left| 1_{\times }\right\rangle \left\langle 1_{\times
}\right| p_{a}\rangle \left\langle p_{a}\right| =\frac{1}{(1+\kappa )}\frac{1%
}{\pi ^{1/2}}\exp \left[ -p_{a}^{2}\right] \left\{ 1+\sin \left[
2\left\langle \alpha \right\rangle p_{a}\right] \right\} .  \label{cat}
\end{equation}
\narrowtext
\noindent Therefore, when Bob builds up the probability distribution $%
p_{1\times }(p_{a})$ of getting outcome $p_{a}$ upon measurement of $\hat{p}%
_{a}$, the distribution should have interference fringes with a period of $%
\pi /\left\langle \alpha \right\rangle $ in the absence of eavesdropping 
\cite{Yurke,Walls}.

To eavesdrop, Eve can, in principle, use a symmetric strategy by applying a
joint unitary operation similar to the one shown in Eqs. (\ref{Sym1}) and (%
\ref{Sym2}). It must involve complex multi-photon interaction between the
single-mode field of the signal states and the probe system, and a physical
mechanism that would enable such an operation has been unknown. Even if such
an operation is realized, we can safely conclude that our proposed scheme is
as secure as the single-photon case as far as this strategy is concerned by
an argument similar to the single-photon case. This conclusion is closely
related to the fact that the quantum mechanical superposition of
macroscopically distinguishable states cannot be noninvasively measured \cite
{Ballentine,Peres}, which is essentially a direct consequence of the quantum
mechanical complementarity. Moreover, this scheme is secure against a
conclusive-measurement attack because the two mutually conjugate sets $%
\left| i_{+}\right\rangle $ and $\left| i_{\times }\right\rangle $ are
linearly dependent. In the rest of the paper, we thus consider only a simple
strategy that can only be used for cryptographic schemes using multi-photon
states, that is, a beamsplitter attack. We show that the intentional
eavesdropping activity will be detected by the legitimate users, and explain
how the eavesdropping is detected.

We consider the following scenario. Eve uses a beam splitter (BS) to sample
part of the signal. She sends Bob the part of the signal transmitted through
the BS and measures the reflected part to gain information about the signal.
What we want to know is how much she can learn and how much she disturbs the
signal state. For this purpose, it is sufficient to calculate Eve's error
rate $Q_{e}^{E}$ on the sifted key for this particular scheme. If we denote
the signal mode defined by the quantum channel as $a$ and an auxiliary mode
introduced at the BS as $b$, the associated joint unitary operation of the
BS on coherent state input is

\begin{mathletters}
\begin{eqnarray}
\left| 0_{+}\right\rangle _{a}\left| 0\right\rangle _{b} &\rightarrow
&U_{BS}\left| \alpha \right\rangle _{a}\left| 0\right\rangle _{b}=\left| 
\sqrt{T}\alpha \right\rangle _{a}\left| -\sqrt{R}\alpha \right\rangle _{b},
\\
\left| 1_{+}\right\rangle _{a}\left| 0\right\rangle _{b} &\rightarrow
&U_{BS}\left| -\alpha \right\rangle _{a}\left| 0\right\rangle _{b}=\left| -%
\sqrt{T}\alpha \right\rangle _{a}\left| \sqrt{R}\alpha \right\rangle _{b},
\end{eqnarray}
where $T=\sqrt{1-R^{2}}$ is the transmission coefficient of the BS\cite
{Barnett}. On the other hand, the same unitary operation transforms the $%
\left| 1_{\times }\right\rangle $ state as \widetext

\end{mathletters}
\begin{equation}
\left| 0_{\times }\right\rangle _{a}\left| 0\right\rangle _{b}\rightarrow
U_{BS}\frac{\left| \alpha \right\rangle _{a}+\left| -\alpha \right\rangle
_{a}}{\sqrt{2(1+\kappa )}}\left| 0\right\rangle _{b}=\frac{1}{\sqrt{%
2(1+\kappa )}}\left\{ \left| \sqrt{T}\alpha \right\rangle _{a}\left| -\sqrt{R%
}\alpha \right\rangle _{b}\right. \left. +\left| -\sqrt{T}\alpha
\right\rangle _{a}\left| \sqrt{R}\alpha \right\rangle _{b}\right\} .
\label{CatEnt}
\end{equation}
\narrowtext
\noindent This indicates that the resultant state is entangled with respect
to modes $a$ and $b$ even though the BS is a linear device. Therefore, noise
is inevitably introduced into the transmission of the $\left| 1_{\times
}\right\rangle $ state. The associated marginal density matrices, $\rho
_{i+}^{B}$ and $\rho _{1\times }^{B}$ for Bob and $\rho _{i+}^{E}$ and $\rho
_{1\times }^{E}$ for Eve after the beamsplitter are calculated as 
\widetext

\begin{eqnarray}
\rho _{i+}^{B} &=&tr_{E}U_{BS}\rho _{i+}^{a}\otimes \left| 0\right\rangle
_{b}\left\langle 0\right| U_{BS}^{-1}=\left| \pm \sqrt{T}\alpha
\right\rangle _{a}\left\langle \pm \sqrt{T}\alpha \right| ,  \label{CatBob1}
\\
\rho _{1\times }^{B} &=&tr_{E}U_{BS}\rho _{1\times }^{a}\otimes \left|
0\right\rangle _{b}\left\langle 0\right| U_{BS}^{-1}  \nonumber \\
&=&\frac{1}{2(1+\kappa )}\left\{ \left| \sqrt{T}\alpha \right\rangle
_{a}\left\langle \sqrt{T}\alpha \right| +\left| -\sqrt{T}\alpha
\right\rangle _{a}\left\langle -\sqrt{T}\alpha \right| \right.  \nonumber \\
&&\left. +{\cal V}_{B}\left( \left| \sqrt{T}\alpha \right\rangle
_{a}\left\langle -\sqrt{T}\alpha \right| +\left| -\sqrt{T}\alpha
\right\rangle _{a}\left\langle \sqrt{T}\alpha \right| \right) \right\} ,
\label{CatBob2}
\end{eqnarray}

\begin{eqnarray}
\rho _{i+}^{E} &=&tr_{B}U_{BS}\rho _{i+}^{a}\otimes \left| 0\right\rangle
_{b}\left\langle 0\right| U_{BS}^{-1}=\left| \mp \sqrt{R}\alpha
\right\rangle _{a}\left\langle \mp \sqrt{R}\alpha \right| ,  \label{CatEve1}
\\
\rho _{1\times }^{E} &=&tr_{B}U_{BS}\rho _{1\times }^{a}\otimes \left|
0\right\rangle _{b}\left\langle 0\right| U_{BS}^{-1}  \nonumber \\
&=&\frac{1}{2(1+\kappa )}\left\{ \left| \sqrt{R}\alpha \right\rangle
_{a}\left\langle \sqrt{R}\alpha \right| +\left| -\sqrt{R}\alpha
\right\rangle _{a}\left\langle -\sqrt{R}\alpha \right| \right.  \nonumber \\
&&\left. +{\cal V}_{E}\left( \left| \sqrt{R}\alpha \right\rangle
_{a}\left\langle -\sqrt{R}\alpha \right| +\left| -\sqrt{R}\alpha
\right\rangle _{a}\left\langle \sqrt{R}\alpha \right| \right) \right\} ,
\label{CatEve2}
\end{eqnarray}
\narrowtext
\noindent where $\rho _{i+}^{a}=\left| i_{+}\right\rangle _{a}\left\langle
i_{+}\right| =\left| \pm \alpha \right\rangle _{a}\left\langle \pm \alpha
\right| $, $\rho _{1\times }^{a}=\left| 1_{\times }\right\rangle
_{a}\left\langle 1_{\times }\right| $, ${\cal V}_{B}=\left| \left\langle 
\sqrt{R}\alpha \right| -\sqrt{R}\alpha \rangle \right| =e^{-2(1-T)\left|
\alpha \right| ^{2}}$, ${\cal V}_{E}=\left| \left\langle \sqrt{T}\alpha
\right| -\sqrt{T}\alpha \rangle \right| =e^{-2T\left| \alpha \right| ^{2}}$
(note that ${\cal V}_{B}{\cal V}_{E}=\kappa $), and the upper sign (resp.
lower sign) corresponds to $i=1$ (resp. $i=0$). Provided that Eve uses an
optimum decision strategy that results in the smallest possible error when
distinguishing two non-orthogonal coherent states $\left| \sqrt{R}\alpha
\right\rangle _{a}$ and $\left| -\sqrt{R}\alpha \right\rangle _{a}$, her
error rate $Q_{e}^{E}$ is given by

\begin{equation}
Q_{e}^{E}=\frac{1-\sqrt{1-{\cal V}_{B}^{2}}}{2}.  \label{CatEveE}
\end{equation}
Such an optimum decision strategy can, in principle, be realized\cite
{Sasaki1,Sasaki2,Sasaki3}.

What Alice and Bob want to do is to evaluate $Q_{e}^{E}$ or Eve's average
information gain $I_{AE}=1-H(Q_{e}^{E})$ as a function of disturbance
observable in the signal that Bob recorded. When we note that Eq. (\ref
{CatBob2}) is formally isomorphic to Eq. (\ref{Vis2}), we find that the most
appropriate measure of the disturbance is the fringe visibility observable
in the probability distribution $p_{1\times }(p_{a})$. From Eq. (\ref
{CatBob2}), $p_{1\times }(p_{a})$ in the presence of eavesdropping can be
easily calculated as \widetext

\begin{equation}
p_{1\times }(p_{a})=Tr\rho _{1\times }^{B}\left| p_{a}\right\rangle
\left\langle p_{a}\right| =\frac{1}{(1+\kappa )}\frac{1}{\pi ^{1/2}}\exp %
\left[ -p_{a}^{2}\right] \left\{ 1+{\cal V}_{B}\sin \left[ 2\sqrt{T}%
\left\langle \alpha \right\rangle p_{a}\right] \right\} .  \label{Pobs}
\end{equation}
\narrowtext
\noindent The fringe visibility is therefore given by ${\cal V}_{B}$. Figure
4 shows Eve's average information gain $I_{AE}=1-H(Q_{e}^{E})$ calculated
from Eq. (\ref{CatEveE}) as a function of the fringe visibility ${\cal V}%
_{B} $. This figure clearly indicates that the amount of information leakage
to eavesdroppers can be estimated from the visibility of the interference
fringe in the probability distribution $p_{1\times }(p_{a})$ of getting
outcome $p_{a}$ upon homodyne-detection measurement of $\hat{p}_{a}$. It is
immediately confirmed that the sum of the squared measures of disturbance $%
{\cal V}_{B}$ and distinguishability ${\cal D}_{B}=1-2Q_{e}^{E}=\sqrt{1-%
{\cal V}_{B}^{2}}$ reaches its expected upper bound of unity; ${\cal D}%
_{B}^{2}+{\cal V}_{B}^{2}=1$. This indicates that the information leakage to
eavesdroppers for a beamsplitter attack reaches its upper bound as does that
for the more sophisticated symmetric eavesdropping strategy. This scheme is
thus secure even against the beamsplitter attack even though the
multi-photon states are used as a signal carrier.

On the other hand, Bob's information gain $I_{AB}=1-H(Q_{e}^{B})$ is easily
evaluated by publicly revealing a part of his sifted keys. In the case of
single-photon implementation, $I_{AB}$ is expected to be unity for this type
of asymmetric attack. Figure 4 compares Bob's information gain $%
I_{AB}=1-H(Q_{e}^{B})$ on the sifted key in the presence of eavesdroppers
for the average photon number $\left| \alpha \right| ^{2}=1$ and $2$ under
the assumption that he performed homodyne detection followed by the standard
decision strategy. In contrast to the single-photon implementation, Bob's
information vanishes in the low fringe-visibility region. This is because
the beamsplitter directs the signal light to Eve and the intensity of the
signal going to Bob falls to zero.

Figure 4 indicates that to learn about Alice's state with some degree of
accuracy, Bob's visibility ${\cal V}_{B}$ must not be too large, which
implies that the reflection coefficient $1-T$ must not be too small. The
requirements for a large information gain and little disturbance are thus
incompatible. A large information gain requires a small transmission
coefficient, while a small disturbance requires a transmission coefficient
close to one, and there is no overlap in the permitted ranges. Therefore,
with this QKD\ scheme, Eve cannot use beamsplitter attack and diverts enough
light to gain any useful information without producing a detectable
disturbance. This confirms the impossibility of noninvasive measurement of
the quantum-mechanical superposition of macroscopically distinguishable
states. The problem for Eve is the vacuum noise from the vacant port of the
BS. If she samples only a small part of the signal, to minimize the
disturbance, the noise from the vacuum state obscures the information
carried by the signal state\cite{Hillery}.

\section{DISCUSSION}

\label{SEC4}

The previous section discussed the ideal situation in which channel loss and
detector inefficiency can be ignored. In this case, the present scheme
enables us to exactly determine the upper bound of the information leakage
to eavesdroppers $I_{AE}$ from the fringe visibility of the probability
distribution $p_{1\times }(p_{a})$. As a result, its security will, in
principle, be provable in the ideal case under the assumption that Alice
does send cat states. This would be the principal advantage of the scheme
over the conventional WCP scheme. This scheme also offers advantages. First,
it involves only quadrature phase measurements, which can be done more
efficiently than photon counting. Second, this scheme can use a more intense
pulse. These advantages may allow us to improve the transmission efficiency
compared to that of the conventional WCP scheme.

However, in the presence of channel loss and detection inefficiency, the
above results need to be reconsidered. It is known that the cat state is so
fragile that the loss of a single photon may easily destroy the interference
fringe observed in the probability distribution $p_{1\times }(p_{a})$.
Moreover, the decoherence rate of the cat state is proportional to the
distance between the two distinguishable coherent states; i.e., it is
proportional to $\sqrt{1-\kappa ^{2}}\sim \left| \alpha \right| $\cite
{Yurke,Walls}. A cat state with a very large average photon number $\left|
\alpha \right| ^{2}$ decoheres rapidly, and thus the present scheme will not
be practical. We here briefly describe the effects of channel loss and
detection inefficiency on the present scheme. Channel loss and detection
inefficiency are, as usual, modeled by a beam splitter that mixes the signal
mode with a vacuum field in an auxiliary mode. Thus, these effects are
essentially analogous to the beamsplitter attack. In other words, neither of
these effects can be distinguished from eavesdropping by means of local
measurement by legitimate users. This would be a weak point in common with
the cryptographic scheme using a multi-photon nonclassical state as a signal
carrier. Let us assume that the overall channel loss is $\zeta $ and the
detector efficiency is $\eta $. The above model then reveals that Eq. (\ref
{Pobs}) is correct if we replace $T\rightarrow \varepsilon \eta T$ in ${\cal %
V}_{B}$ and Eq. (\ref{Pobs}). This means that ${\cal V}_{B}$ is reduced by a
factor $e^{-2(1-\varepsilon \eta )T\left| \alpha \right| ^{2}}$ and $0\leq 
{\cal V}_{B}\leq e^{-2(1-\varepsilon \eta )\left| \alpha \right| ^{2}}$.
Thus, unless $\varepsilon \eta \neq 1$, the visibility is less than unity
even if $T=1$; i.e., no eavesdropper is present. The legitimate users can
measure the detection efficiency locally though. Thus, its effect may be
subtracted when determining the upper bound of the information leakage to
Eve from the observed visibility. Moreover, if legitimate users have a
reliable way to evaluate the channel loss, its effect may also be
subtracted. However, when $2(1-\varepsilon \eta )\left| \alpha \right|
^{2}\gg 0$, ${\cal V}_{B}\approx 0$. Eavesdropping would then be very
difficult to detect from the subtle changes in ${\cal V}_{B}$, and this
scheme would not be practical. Thus, $1-\varepsilon \eta $ and $\left|
\alpha \right| ^{2}$ need to be very small for this scheme to work. In
practice, even a minor 3-dB loss ($\varepsilon \eta =0.5$) will make it hard
to use this scheme, unless $\left| \alpha \right| ^{2}$ is small. It is this
extreme sensitivity of the nonclassical field state to the environment that
enables us to detect eavesdropping. In contrast, the channel loss is simply
discarded in the WCP scheme, but this discarding also makes the WCP scheme
vulnerable because eavesdroppers have chance to use it while substituting a
superior channel to escape detection.

On the other hand, there is a lower bound on $\left| \alpha \right| ^{2}$
that enables use of the cat state to detect eavesdropping. To evaluate the
fringe visibility, there should be at least one oscillation in the
distribution $p_{1\times }(p_{a})$ within the Gaussian contour $\exp \left[
-p_{a}^{2}\right] $. This requirement should impose the inequality $\Delta
p_{a}=\pi /(2\sqrt{2\varepsilon \eta T}\left| \alpha \right| )<2\sqrt{\ln 2}$%
. If we note that $\varepsilon \eta T\leq 1$, $\left| \alpha \right| >\pi /(4%
\sqrt{2\ln 2})\sim 0.67$ is required. Thus, a cat state with an average
photon number of the order of unity is appropriate for our scheme. In this
sense, what is needed is not a ``macroscopic'' quantum superposition but a
``mesoscopic'' quantum superposition which should be easier to create. The
present scheme is effective only if a good channel with low loss, a highly
efficient detector, and a mesoscopic cat state are available.

Hillery's idea of using phase-sensitive amplifiers to boost the signal and
partially compensate for the effect of losses for a cryptographic scheme
using the squeezed state is interesting. When a device that can amplify the
cat state becomes available in the future, his idea can also be applied to
the present scheme: if such a device is used in the secure station in the
channel, it can partially compensate for the effect of losses by amplifying
only the cat state in accord with the state Alice sends, and otherwise it
can compensate the effect by randomly amplifying the signal and discarding a
part of the data.\cite{Hillery}

The problem with the cryptographic scheme using a multi-photon nonclassical
state comes from the fact that the state after eavesdropping and the state
after losses are indistinguishable. A cryptographic scheme using a single
photon seems not to have such a problem. This is because the state after
photon loss is a vacuum state that spans a subspace different from the
signal space of a single-photon state. Such a scheme can conclusively
distinguish a photon-loss event and uses only the photons that {\it did not}
get absorbed. Thus, it can overcome 10 dB of losses. That, in essence, is
why conventional quantum cryptography using single photons can work so well
in the presence of channel loss and detection inefficiency. We should,
however, remember that this is true only with the assumption that the
apparatus used by Alice to produce the photon is perfect\cite{Mayers1}. To
confirm that this assumption is valid, we must check that the signal space
is limited to two-dimensional space; i.e., whether the photon source is a
good single-photon source. This check can be accomplished through
observation of the nonlocal properties inherent in quantum mechanics, such
as violation of Bell's inequality\cite{Mayers2} and Bell's theorem without
inequalities associated with an entangled tripartite system,\cite
{GHZ,Mayers3} to ensure that the photon source is a perfect single-photon
source. This is because no local photon source other than a perfect
single-photon source can reproduce the results observed nonlocally between
legitimate users.\cite{Mayers2,Mayers3} However, note that the rigorous test
for a violation of Bell's inequality and Bell's theorem associated with an
entangled tripartite system is possible only if the product of the channel
loss and detection efficiency is higher than some lower limit of around
80\%. \cite{Clauser,Garg,Larsson} If the loss and efficiency product is
below this lower limit, an auxiliary assumption (a fair-sampling assumption)
that the fraction of detected pairs is representative of the entire ensemble
is required to rule out any local realistic model that can reproduce the
observed results and to prove the security of the BB84 protocol without any
loophole.\cite{Clauser,Garg,Larsson}

The current feasibility of the present scheme is limited by the difficulty
of preparing the cat state with today's technology in addition to channel
loss and detector inefficiency. However, a development of a quantum gate
will help us to obtain the cat state through a swapping operation\cite
{Aharonov} between a coherent state and a more easily created superposition
state of a single quanta\cite{Cochrane,Gerry}.

\section{CONCLUSION}

\label{SEC5}

It is a distinct feature of the ideal BB84 protocol that its security is
mathematically provable. This is possible because quantum mechanical
complementarity enables us to determine the upper bound of the information
leakage to eavesdroppers. We have shown that complementarity allows also use
of multi-photon states as a signal carrier in the BB84 QKD protocol, and
have described a scheme that uses two nearly orthogonal coherent states to
carry the key where the superposition of these states protects the
communication channel from eavesdropping. This scheme is based on
complementarity as is the conventional BB84 scheme. We expect this scheme to
be as secure as the conventional single-photon scheme and secure against any
eavesdropping strategy. The disappearance of interference fringes in the
homodyne detection used to decode the key clearly indicates eavesdropping
activity, and the upper bound of information leakage to eavesdroppers can be
exactly determined from the visibility of the interference fringes which is
measurable from the homodyne detection. As a result, this scheme will be
provably secure as long as a good channel with negligible loss, a highly
efficient detector, and a mesoscopic cat state are available.

Unfortunately, this scheme is very sensitive to losses and is not practical
in the presence of high channel loss and detection inefficiency. In this
case, a single-photon implementation would be preferable, although the
provable security still requires low channel loss and high detection
efficiency.

\acknowledgments

We thank Dominic Mayers for his helpful discussions concerning the validity
of the present scheme.

\begin{figure}[tbp]
\caption{Upper plot: Bob's information gain $I_{AB}$, Eve's information gain 
$I_{AE}$ and their sum $I_{AB}+I_{AE}$ are plotted against Bob's error
probability $Q_{e}^{B}$ when Eve applys an optimum eavesdropping strategy.
Lower plot: measures of information gained by Bob ($G^{B}$) and by Eve ($%
G^{E}$) are plotted.}
\label{F1}
\end{figure}

\begin{figure}[tbp]
\caption{The relevant subspace of the four weak coherent states in the
entire Hilbert space (the Fock space). The parts of the four circles that do
not overlap indicate the linear independence of the states.}
\label{F2}
\end{figure}

\begin{figure}[tbp]
\caption{The basic idea of the proposed QKD scheme. Alice and Bob use two
nearly orthogonal coherent states to carry the key and the superposition of
these states (cat states) to protect from eavesdropping. Eavesdropping is
detected from the disappearance of the interferential fringes in the
distribution of the outcome when a certain quadrature component is measured
by the homodyne detection.}
\label{F3}
\end{figure}

\begin{figure}[tbp]
\caption{ Information leakage to $I_{AE}$ (a solid line) and Bob's average
information gain $I_{AB}$ (broken lines) as a function of the fringe
visibility ${\cal V}_{B}=e^{-2(1-T)\left| \protect\alpha \right| ^{2}}$ in
the probability distribution $p_{1\times }(p_{a})$ recorded by Bob. $I_{AB}$
was evaluated at the average photon number $\left| \protect\alpha \right|
^{2}=2$ and $\left| \protect\alpha \right| ^{2}=1$.}
\label{F4}
\end{figure}

\end{document}